\documentclass[10pt,a4paper]{article}
\usepackage{amsmath}
\usepackage{amssymb}

\def\Tr{\mathop{\rm Tr}\nolimits}

\title{QCD, Wick's Theorem for KdV $\tau$-functions and the String Equation.
}

\author{H. W. Braden$^{1}$\thanks{hwb@ed.ac.uk},
A.Mironov$^{2,3}$\thanks{mironov@lpi.ac.ru, mironov@itep.ru},
A.Morozov$^{3}$\thanks{morozov@vx.itep.ru} \
\\ \normalsize \em $^{1}$Department of Mathematics and
Statistics,  The University of Edinburgh,  Edinburgh, UK
\\
\normalsize \em $^{2}$ Theory
Department, Lebedev Physics Institute, Moscow
~117924, Russia
\\
\normalsize \em $^{3}$ITEP, Moscow
117259, Russia }

\date{}

\begin{document}

\renewcommand{\thepage}{}
%\begin{titlepage}

\maketitle

\vspace{-7.7cm}

\begin{center}
\hfill EMPG-01-07\\
\hfill FIAN/TD-09/01\\
\hfill ITEP/TH-23/01\\
\hfill hep-th/0105169
\end{center}

\vspace{5.0cm}

\begin{abstract}
Two consistency conditions for partition functions established
by Akemann and Dam-gaard in their studies of the fermionic mass dependence of
the QCD partition function at low energy ({\it a la}
Leutwiller-Smilga-Verbaarschot) are
interpreted in terms of integrable hierarchies. Their algebraic relation is
shown to be a consequence of Wick's theorem for $2d$ fermionic correlators
(Hirota identities) in the special case of the 2-reductions of the KP hierarchy
(that is KdV/mKdV). The consistency condition involving derivatives
is an incarnation of the string equation associated with the particular matrix
model (the particular kind of the Kac-Schwarz operator).
\end{abstract}

\paragraph{1.}
Recently G.Akemann and P.Damgaard \cite{AK} established two consistency
relations for the partition functions of matrix models which connect
finite-volume partition functions with different fermion numbers.
In the course of their paper the similarity of these relations with
different bilinear relations for the $\tau$-functions of the KP hierarchy
were noted and they conjectured they were in fact related.
This letter provides a comment on the origin of these reduction formulae.
We show they are direct
corollaries of Wick's theorem and the string equations of the $\tau$-function
theory.  Application of this theorem is possible because the partition
functions of Cartanian matrix models are KP/Toda $\tau$-functions (see
\cite{UFN, dido} and references therein, especially \cite{GMMMO} and
\cite{GKM}, for the generic theory and \cite{KMMM,MMS} for applications to
unitary-matrix models).  The algebraic relation (``Consistency Condition I''
of ref.\cite{AK}) is true for {\it any} 2-reduction of KP $\tau$-function,
which are essentially the KdV or mKdV reductions\footnote{The KdV and mKdV
hierarchies are basically the same reduction of the (m)KP hierarchy. One
deals with the mKdV case when one additionally considers the discrete (Toda)
time ``evolution". While generically the evolution of the KP $\tau$-function
$\tau_n$ in this discrete time is governed by the Toda dynamics, in the case
of the 2-reduction there exist only two different values of the zero time
due to the periodic boundary conditions, $\tau_{n+2}=\tau_n$. The
corresponding $\tau$-functions $\tau_0$ and $\tau_1$ give the mKdV
hierarchy.}, of which the Brezin-\-Gross-\-Witten
(BGW) model \cite{BGN} under actual examination in \cite{AK} is an example.
The differential relation (``Consistency Condition II'') is the string equation,
depending on the specifics of the particular model. (The BGW model is studied
from the perspective of integrability theory in ref.\cite{MMS}.)

These consistency conditions are imposed onto the QCD partition function as a
function of fermionic masses $\mu_i$ at very low energies, which is of the
general form
\begin{equation}
{\cal Z}_\nu^{(N_f)} (\{\mu\}) ~=~ \det \Phi(\{\mu\})/\Delta(\{\mu^2\}).
\label{ZUE}
\end{equation}
Here the matrix $\Phi$ is defined by
\begin{equation}
\Phi(\{\mu\})_{ij}=\phi_i(\mu_i),~~~~i,j=1,\ldots,N_f,
\label{Adef}
\end{equation}
for some functions $\phi_i(\mu)$ yet to be specified.
The denominator is given by the Vandermonde determinant of the squared masses
\begin{equation}
\Delta (\{\mu^2\})\ \equiv\ \prod_{i>j}^{N_f}(\mu_i^2-\mu_j^2)
\ =\ \det_{i,j}\left[ (\mu_i^2)^{j-1}\right] .
\label{Vandermonde}
\end{equation}

Now the consistency conditions I and II of \cite{AK} acquire the forms
respectively
\begin{equation}\label{I}
\det_{1\leq a,b\leq k}\left[
\frac{{\cal Z}_{\nu}^{(N_{f}+2)}(\{\mu\},\xi_a,\eta_b)}
{{\cal Z}_{\nu}^{(N_{f})}(\{\mu\})}\right]
=
\prod_{i<j}^k(\xi_i^2-\xi_j^2)(\eta_i^2-\eta_j^2) ~
\frac{{\cal Z}_{\nu}^{(N_{f}+2k)}
(\{\mu\},\xi_1,\ldots,\xi_k,\eta_1,\ldots,\eta_k)}
{{\cal Z}_{\nu}^{(N_{f})}(\{\mu\})}  ,
\end{equation}
and
\begin{equation}\label{II}
\begin{split}
{\cal Z}_\nu^{(N_f+2)} (\{\mu\},\xi,\eta) &=
\frac{1}{(\xi^2-\eta^2){\cal Z}_\nu^{(N_f)} (\{\mu\})}
\times\\
&  \left[ \left( (\sum_{i=1}^{N_f}\mu_i\partial_{\mu_i}
+\xi\partial_{\xi})
{\cal Z}_\nu^{(N_f+1)}(\{\mu\},\xi)\right)
{\cal Z}_\nu^{(N_f+1)}(\{\mu\},\eta) ~-~
\left(\xi \leftrightarrow \eta\right)\right].
\end{split}
\end{equation}
In what follows we discuss these equations, their origins and
solutions.

\paragraph{2.} We begin with the universal
(in the world of 2-reductions of KP) algebraic relation (\ref{I}).
It states, in particular, that for {\it certain sets} of functions
$\phi_i(\zeta)$, $i = 1,2,\ldots$, and for any $k$

\begin{equation}
\frac{\det^{(2k)}_{(i,j)} \phi_i(\zeta_j)}{\Delta(\zeta^2)} \ = \
\frac{1}{\Delta(\xi^2)\Delta(\eta^2)}\
{\det}^{(k)}_{(a,b)}\left(\frac{\det^{(2)}
\left|\begin{array}{cc}
\phi_1(\xi_a)&\phi_1(\eta_b)\\\phi_2(\xi_a)&\phi_2(\eta_b)
\end{array}\right|}{\xi_a^2-\eta_b^2}\right).
\label{ID}
\end{equation}
Here $i,j = 1,\ldots,2k$, while $a,b = 1,\ldots,k$ and the set of $2k$ variables
$\{\zeta_j\}$ is (arbitrarily) split into two subsets $\{\xi_a\}$ and $\{\eta_b\}$
of $k$ variables each
(for example, $\zeta_a = \xi_a$, $\zeta_{a+k} = \eta_a$).

Obviously, relation (\ref{ID}) implies that all the functions $\phi_i(\zeta)$
are expressed in a certain way through the first two, $\phi_1(\zeta)$ and
$\phi_2(\zeta)$.
{\it Sufficient} conditions for (\ref{ID}) to be true are, for example,
the recurrence relations:
\begin{equation}
\phi_{i+2}(\zeta) = \zeta^2\phi_i(\zeta) + \sum_{j=1}^{i+1}A_{ij}\phi_j(\zeta),
\ \ A_{ij} = const.
\label{KdV}
\end{equation}
In the $\tau$-function theory this recursion corresponds to the (m)KdV
reduction.
The derivation of (\ref{ID}) from (\ref{KdV}) is an easy algebraic
exercise (it can be done straightforwardly by induction in $k$ or one can
expand $\phi_i(\zeta) = \zeta^{i-1}\sum_{k=0}^\infty p_{ik}\zeta^{-k}$ as a
formal series in $\zeta^{-1}$ and express the determinants through the
characters of $SL(N)$, see refs.\cite{KMMM,MMS} for examples).

Observe that, among other equations, the consistency condition I (\ref{I})
and similarly (\ref{ID}) are invariant with respect to  multiplication of
all the $\phi_i(\mu)$ by an arbitrary function $f(\mu)$.

\paragraph{3.} We shall now comment on how (\ref{ID}) and, more
generally, (\ref{I}), follow from the $\tau$-function theory.  There are three
ingredients needed from that theory:
\begin{enumerate}
\item
Wick's theorem and its realisation (the Hirota equation) in terms of the KP/Toda
$\tau$-function in Miwa coordinates (involving positive and negative shifts
of time-variables, see below);

\item
Determinant formulae for $\tau$ in Miwa coordinates (with positive shifts only);
\item
The fact that $\tau$ for a $p$-reduced hierarchy is essentially independent
of the times $t_{pk}$. Specifically, the (m)KdV $\tau$-function is
independent of even times:
\begin{equation}
\partial \tau_{KdV}/\partial t_{2n} = 0.
\label{evenT}
\end{equation}
Generally \cite{GKM,dido} the $\tau$-function is only defined up to times
linear in the
exponential: ${\partial^2 \log\tau}/{\partial t_{pk}\sp2}=0.$ This is
equivalent to the freedom to multiply the functions $\phi_i(\mu)$ by a function
$f(\mu)$ mentioned earlier. Here we are making the specific choice that
yields (\ref{evenT}).
\end{enumerate}

This last property allows one to essentially identify the positive and negative
Miwa shifts of time-variables for the (m)KdV $\tau$-functions and combine
Wick's theorem and the determinant representation into a common identity
(\ref{ID}) as we shall now demonstrate.

\paragraph{4.} Wick's theorem (in the restricted form which we need here)
is an identity between correlators for any free-fermion theory with action
of the form $\int \psi(\xi) G(\xi,\eta) \tilde\psi(\eta) d^D\xi d^D\eta$.
It says
\begin{equation}
\langle\prod_{a=1}^k \psi(\xi_a)\tilde\psi(\eta_a)\rangle \ =
\ {\det}^{(k)}_{(a,b)} \langle\psi(\xi_a)\tilde\psi(\eta_b)\rangle.
\label{WTh}
\end{equation}
For the KP/Toda $\tau$-functions, when $D=2$ and $G(\xi,\eta)$ is a
first-order differential operator, $G(\xi,\eta) =
\delta(\xi,\eta)(\bar\partial_\xi + \bar A(\xi))$, the correlators can be
written as
\begin{equation}
\langle\prod_{a=1}^k \psi(\xi_a)\tilde\psi(\eta_a)\rangle \ =
\frac{\tau_{KP}\left( t + \sum_a([\xi_a] - [\eta_a])\right)}{\tau_{KP}(t)}\cdot
\frac{\Delta(\xi)\Delta(\eta)}{\prod_{a,b}(\xi_a - \eta_b)}.
\label{FCvsTAU}
\end{equation}
The dependence on times $\{t\}$ on the left-hand-side of this expression
is hidden in the definition of the brackets.
Here the arguments of the $\tau$-function are Miwa-shifted time-variables
\begin{equation}
t_n + \frac{1}{n}\sum_{a=1}^k (\xi_a^{-n} - \eta_a^{-n})
\label{Miwa}
\end{equation}
and the Vandermonde
determinants ($\Delta(\xi) = \prod_{a<b}^k(\xi_a-\xi_b)$ etc.) arise from the
normal orderings. While
$$
\tau_{KP}\left( t + \sum_a([\xi_a] - [\eta_a])\right) = \tau_{KP}(t)
\langle\ :\prod_{a=1}^k \psi(\xi_a)\tilde\psi(\eta_a): \rangle
$$
is non-singular at coincident points $\xi$ and $\eta$, the correlator in
(\ref{FCvsTAU}) vanishes when any two of $\xi$ or any two of $\eta$ collide
and have a pole when any $\xi$ approaches any $\eta$. The Vandermonde
determinants encode this singular behaviour and corresponds to the naive
(``Riemann sphere'') normal ordering.
One sometimes uses more sophisticated orderings associated with
non-trivial Riemann surfaces, but this is not the case (for the matrix models)
under consideration.

\paragraph{5.} In Wick's theorem, the numbers of $\xi$ and $\eta$ variables
(i.e. of the $\psi$ and $\tilde\psi$ operators, or of the positive and
negative Miwa shifts) are necessarily the same.  However, one can take a limit
when all the $\eta$ points collide, say at infinity.  Then Wick's theorem
implies a peculiar determinant formula for the $\tau$-function in Miwa
parametrisation \cite{GKM} (with only positive Miwa shifts remaining):
\begin{equation}
\frac{\tau_{KP}( t + \sum_a[\xi_a])}{\tau_{KP}(t) } =
\frac{{\det}^{(k)}_{(a,b)} \phi_a(\xi_b)}{\Delta(\xi)}.
\label{DF}
\end{equation}
From this perspective, the functions $\phi_i(\xi)$ arise as expansion
coefficients in
$$
<\psi(\xi) \tilde\psi(\eta)> = \sum_{i=0}^\infty \phi_i(\xi)\eta^{-i}.
$$
Note that all the time-dependence is hidden in the shapes of the
functions $\phi_i$. In particular, one can make a further Miwa transformation
of the times $t$ in (\ref{DF}) enlarging the right-hand determinant of
(\ref{DF}) and correspondingly changing the functions $\phi_i$. The same is
true of the $\mu$-dependency in formulae (\ref{I})-(\ref{II}): one
could hide the dependence on $\{\mu_i\}$ in the functions
$\phi_j$, though they can of course be taken out to produce determinant
formulae with additional sets of $N_f$ parameters $\mu_1,\ldots,\mu_{N_f}$.

At this stage nothing is gained by substituting the determinant formula
(\ref{DF}) back into the Wick identity (\ref{WTh}), from which it was
derived: expansion of both sides of (\ref{WTh}) in powers of $\eta_a^{-i_a}$
will produce nothing new.

\paragraph{6.} However, in the case of (m)KdV $\tau$-functions there is
an amusing way of introducing (another set of) $\eta$ parameters back into the
game. Namely, because of property (\ref{evenT}),
\begin{equation}
\tau_{KdV} ( t + [\xi] - [-\eta]) = \tau_{KdV}(t + [\xi] + [\eta])
\label{KDVid}
\end{equation}
since the difference between the arguments is seen (by (\ref{Miwa})) only in
the even times $t_{2n}$, on which $\tau_{KdV}$ does not depend.
Now one can apply Wick's theorem (\ref{WTh}) with the substitution
(\ref{FCvsTAU}) to the left-hand-side of (\ref{KDVid}) and the determinant
formula (\ref{DF}) to its right-hand-side.
Straightforward calculation produces the relation (\ref{ID}).

Note now that for the ratio of determinants (\ref{DF}) to be a
$\tau$-function, the functions $\phi_i(\mu)$ are to have prescribed
asymptotics at large $\mu$, usually $\phi_i(\mu)\sim \mu^{i-1}$ (see, however,
\cite{KMMM}). This ensures that the $\tau$-function is an ($N_f$-independent)
function of the
usual times $t_k=1/k\sum_i^{N_f}\mu_i^{-k}$, and so allows
an expansion as a formal series in the time variables $t_k$. These
asymptotic properties remain unchanged when the functions $\phi_i(\mu)$ are
multiplied by a function $f(\mu)$ that behaves like $1$ for large $\mu$.
In particular, if $\log f(\mu)$ expands into a
Taylor series in $1/\mu$, such a redefined $\tau$-function is multiplied
by a linear exponential of times $t_k$. Therefore, the property
(\ref{evenT}) holds for the (m)KdV $\tau$-function only for a very specific
normalisation of the $\phi_i(\mu)$.

We have now shown how consistency condition I (\ref{I}) can be understood
in terms of Wick's theorem for a particular (m)KdV $\tau$-function.
It remains to discuss consistency condition II and the matrix theory origins.

\paragraph{7.}
While the algebraic relation (\ref{I}) allows one to impose restrictions on
$\phi_i(\mu)$ that lead to an (m)KdV $\tau$-function, i.e. express all
$\phi_i(\mu)$'s through the first two, the differential relation
(\ref{II}) expresses all of the $\phi_i(\mu)$'s in terms of derivatives of the
first.  The combination of these two restrictions then fixes the $\phi_i(\mu)$
(up to a linear combination of rows in the determinant and a trivial prefactor).
In terms of integrable hierarchies, this relation is called
the string equation \cite{GKM}. The role of this equation is to fix a unique
solution of the integrable hierarchy (specified by the algebraic relation
(\ref{I}) and the form of $Z_\nu^{(N_f)}$ (\ref{ZUE})).

Let us start with the simplest case of $N_f=0$. Then, (\ref{II}) implies
that\footnote{Up to a term proportional to
$\phi_1(\mu)$; there is always a freedom to add to any $\phi_k(x)$ a linear
combination of $\phi_i(\mu)$ with $i<k$ with constant coefficients, since
it does not change the determinant in (\ref{DF}). Hereafter, we denote this
freedom with the sign ``$\sim$".}
$\phi_2(\mu)\sim\mu\partial\phi_1(\mu)/\partial \mu$. This is, however, still not a
restriction on $\phi_1(\mu)$ itself.
In order to see how (\ref{II}) restricts all the $\phi_i(\mu)$,
one needs to consider it at all values of $N_f$.
In doing so, it is necessary to use the
equation that is {\it tautologically satisfied} by the ratio of determinants
(\ref{DF}) \cite{GKM}\footnote{The authors of \cite{AK} correctly suggested its
relevance to the problem. However, let us emphasise again that this equation
does not select out any particular $\tau$-function, or any particular $\phi_i(\mu)$.
This is just another form of formula (\ref{DF}).}:
\begin{equation}\label{GKM}
\tau_{N_f+2}(\{\mu\},\xi,\eta)={1\over (\xi -\eta )\tau_{N_f}(\{\mu\})}
\left[\tau_{N_f+1}(\{\mu\},\xi)\hat\tau_{N_f+1}(\{\mu\},\eta)-
(\xi\leftrightarrow\eta)\right]
\end{equation}
where the hat over $\tau_N$ means that the last row $\phi_N(\{\mu\})$ is
substituted with $\phi_{N+1}(\{\mu\})$. This equation is typically used for
derivation of the string equation \cite{GKM,MMS}.

Now, since
$\sum_i\mu_i\partial_{\mu_i}\Delta_{N_f}(\mu^2)=N_f(N_f-1)\Delta_{N_f}(\mu^2)$, one can rewrite (\ref{II}) as
\begin{equation*}
\begin{split}
{\det}^{(N_f+2)}\Phi(\{\mu\},&\xi,\eta)\cdot{\det}^{(N_f)}\Phi(\{\mu\})\\&=
\left[\left(\sum_i\mu_i\partial_{\mu_i}+\xi\partial_{\xi}\right)
{\det}^{(N_f+1)}\Phi(\{\mu\},\xi)\right]{\det}^{(N_f+1)}\Phi(\{\mu\},\eta)-
(\xi\leftrightarrow\eta).
\end{split}
\end{equation*}
Comparing with (\ref{GKM}), we conclude that
\begin{equation}
\begin{split}
\Bigg[\left(\sum_i\mu_i\partial_{\mu_i}+\xi\partial_{\xi}\right)
{\det}^{(N_f+1)}&\Phi(\{\mu\},\xi)\Bigg]
{\det}^{(N_f+1)}\Phi(\{\mu\},\eta)-(\xi\leftrightarrow\eta)\\ &=
{\det}^{(N_f+1)}\hat \Phi(\{\mu\},\xi){\det}\sp{(N_f+1)}\Phi(\{\mu\},\eta)-
(\xi\leftrightarrow\eta).
\end{split}
\end{equation}
with the evident notation $\hat \Phi$ denoting the matrix whose last row
$\phi_N(\{\mu\})$ has been substituted with $\phi_{N+1}(\{\mu\})$.
Induction on $N_f$ (we have already shown it true for $N_f=0$ above) shows
this equation is solved
if
\begin{equation}
{\det} \hat \Phi=\left(\sum_i\mu_i\partial_{\mu_i}\right)\det \Phi +
N_f\alpha\ {\det} \Phi,
\label{stringsol}
\end{equation}
$\alpha$ being an arbitrary constant.
In fact this relation is another version of the string equation (cf. with
\cite{GKM}). The term proportional to ${\det} \Phi$ can always be absorbed
by the simultaneous rescaling all $\phi_i(\mu)$ with $\mu^\alpha$
and (\ref{stringsol}) yields in its turn that
$\phi_i(\mu)\sim(\mu\partial_\mu)^{i-1}\phi_1(\mu)$.

Therefore, the consistency condition II really restricts the functions
$\phi_i(\mu)$. Combining the form
$\phi_i(\mu)\sim(\mu\partial_\mu)^{i-1}\phi_1(\mu)$ together with the reduction
condition (\ref{KdV}) (=consistency condition I), one
arrives at the solution \cite{AK}
\begin{equation}\label{Bessel}
\phi_i(\mu)\sim \mu^{\alpha+i-1}I_{\nu+i-1}(\mu)
\end{equation}
where $I_k(\mu)$ are the modified Bessel functions, $\nu$ and $\alpha$ are
arbitrary parameters (our solution slightly differs from that in \cite{AK}
by the additional parameter $\alpha$).

\paragraph{8.}

The solution (\ref{Bessel}) obtained for $Z_\nu^{(N_f)}$ can also  be
presented in terms of the matrix integral \cite{MMS}
\begin{equation}\label{ZZ}
Z_\nu^{(N_f)}(\{\mu\})\sim\int dXe^{\Tr\left[M^2X/4-(\nu+N_f)\log X+1/X +
(\alpha-\nu)\log M\right]}
\end{equation}
where the integral is taken over Hermitian $N_f\times N_f$ matrix $X$ and
$\mu_i$ are the eigenvalues
of the matrix $M$. For the particular values $\alpha=\nu=0$ this integral may
be rewritten as an integral over unitary $N_f\times N_f$ matrix $U$
\cite{MMS}
\begin{equation}\label{Z}
Z^{(N_f)}_0(\{\mu\})\sim\int dU e^{\Tr\left[J^\dag U + U^\dag J\right]},\ \
\ \ M^2=JJ^\dag.
\end{equation}

The function $Z_\nu^{(N_f)}(\{\mu\})$ by itself is certainly not a
$\tau$-function for the Miwa coordinates (11) we are using here 
(chosen because of the formulation of [1]).  As we noted earlier,
for the ratio of determinants (\ref{DF}) to be a ($\tau$-)function of the times
$t_k$ we required the asymptotics of $\phi_i(\mu)$ at large $\mu$ to be
normalised by $\mu^{i-1}$ \cite{GKM}, which are not those of (\ref{Bessel}).
 An easy calculation shows that, in order to obtain a $\tau$-function, one
needs to include several prefactors \cite{MMS}
\begin{equation}\label{tau}
\tau_{\nu}(\{t\})=\prod_{i<j}(\mu_i+\mu_j)\prod_i \sqrt{\mu_i}\mu_i^{-\alpha }
e^{-\mu_i}Z_\nu^{(N_f)}(\{\mu\}).
\end{equation}
Here the parameter $\nu$ plays role of the discrete time (in the standard
notation of \cite{MMS} it has the opposite sign).
Such prefactors typically emerge in matrix model theory: to obtain
a $\tau$-function, one should factor out a quasi-classical piece from the
matrix integral.  For the particular integral (\ref{ZZ}) this exactly gives
(\ref{tau}), and in fact reference \cite{MMS} also shows the independence of 
even times we have wanted here. Thus we return to the matrix model integrals 
that inspired the consistency conditions we have been discussing.

In concluding let us observe that these integrals can be further generalised
yielding other $\tau$-functions.
In particular, one can consider an arbitrary function $V(X)$ polynomial in
$X$ or $1/X$. Then, the integral
\begin{equation}
C_V\int dXe^{\Tr\left[XL-(\nu+N_f)\log X+V(X)\right]}
\end{equation}
results in a two-dimensional Toda lattice $\tau$-function,
\begin{equation}
\tau_\nu(\{t\})=
\frac{{{\det}_{ij}\left[\mu_i^\beta\psi_{j-\beta}(\mu_i)\right]}}
{\Delta (\mu)}.
\end{equation}
Here $C_V$ is the normalisation factor cancelling the quasi-classical part of
the integral,
and $\mu_i$ are related with the eigenvalues $l_i$ of the matrix $L$ via
the saddle point condition $l+V'(\mu)=0$ or $l+V'(1/\mu)=0$ depending on whether
$V(X)$ is a polynomial of $X$ or $1/X$ respectively. The time variables are
again $t_k=1/k\sum_i\mu_i^{-k}$ and
\begin{equation}
\psi_{i-\beta}(\mu)\equiv {1\over\sqrt{V''(\mu)}}e^{V(\mu)-\mu V'(\mu)}\int
dx x^{(i,\beta)}e^{xl+V(x)}
\end{equation}
where $(i,\beta)$ is equal to $i-\beta-1$ for $V(X)$ being a polynomial in $X$
and $\beta-i$ for a polynomial in $1/X$.

\bigskip

Our work is supported by: the NATO grant PST.CLG.976955 (all the three
authors),
the CRDF grant CRDF \#6531 (A.M.'s),
the RFBR grants  00-02-16101a (A.Mir.) and 01-02-17488 (A.Mor.), INTAS grants
00-334 (A.Mir.) and 00-561 (A.Mor.), and the Russian President's Grant
00-15-99296 (A.Mor).

\end{document}